\documentclass[10pt,aps,prl,twocolumn,superscriptaddress,longbibliography]{revtex4-1}

\usepackage{graphics}
\usepackage{epsfig}
\usepackage{dcolumn}
\usepackage{bm}
\usepackage{amsmath}
\usepackage{dsfont} 
 \usepackage{amsfonts}
\usepackage{latexsym}
\usepackage{amssymb}
\usepackage{color}
\usepackage{hyperref}
\usepackage[normalem]{ulem}

 \newcommand{\tb}[1]{}

\newcommand{\g}[1]{{\bf #1}}

\newcommand{\be}{\begin{equation}}
\newcommand{\ee}{\end{equation}}
\newcommand{\bea}{\begin{eqnarray}}
\newcommand{\eea}{\end{eqnarray}}
\newcommand{\ba}{\begin{eqnarray*}}
\newcommand{\ea}{\end{eqnarray*}}
\newcommand{\dagga}{{\phantom{\dagger}}}

\begin{document}

\title{Rabi-resonant behavior of periodically-driven correlated fermion systems}
 
\author{Marcin P\l{}odzie\'n}
\email{mplodzien@magtop.ifpan.edu.pl}
\author{Marcin M. Wysoki\'nski}
\email{wysokinski@magtop.ifpan.edu.pl}
\affiliation{International Research Centre MagTop, Institute of
  Physics, Polish Academy of Sciences,\\ Aleja Lotnik\'ow 32/46,
  PL-02668 Warsaw, Poland}

\date{\today}

\begin{abstract}
Fermi-Hubbard system with a periodically-modulated interaction has been recently shown to resonantly absorb energy at series of drive frequencies. 
In the present work, with the help of static perturbation theory, we argue that driving couples to Hubbard bands in a similar fashion as a classical field couples to a two-level atom. The latter, significantly simpler set-up, described by the Rabi model, is known to support resonant rapid energy absorption at virtually the same series of driving frequencies. Our interpretation is also supported by the equivalency between dynamics 
of the two-site periodically-driven Hubbard and Rabi models. 
Relying on this insight, we establish that the same Rabi-resonant behavior is displayed also by fermions confined in a harmonic trap with periodically-driven contact interactions.
\end{abstract}
\maketitle
 
{\it Introduction.}
Strongly correlated fermion systems driven far from equilibrium 
can display novel and exotic behavior \cite{Mihailovic2016,Mitrano2016,Giannetti2017,Marsi2017,Fabrizio2017,Esslinger2018b}. 
However, revealing underlying physical principles poses a significant theoretical challenge. Even understanding of the post-quench dynamics of the simplest Fermi-Hubbard model is far from being conclusive \cite{Kollar2009,Schiro2010,Potthoff2016,tSWT},  due to (among others)  limitations of non-equilibrium methods that were used.  

In this light, the perturbative approaches to correlated fermions based on the Floquet theorem  \cite{Eckardt2017,Eckardt_2015,Liberto2014,Itin2015,Polkovnikov2016,Coulthard2017,Mentink2015,Katsnelson2016,Katsnelson2017}, though restricted to time-periodic drivings, have turned out to be very efficient and powerful tools,  capturing well  the experimental realizations \cite{Esslinger2017,Esslinger2018a,Esslinger_arxiv,Esslinger_arxiv2}. 
Usually,  the accurate determination of the energy absorption in an arbitrarily driven correlated fermion system involves heavy numerical simulation \cite{Kollar2009,Werner2011,Werner2014}. However, for periodically-driven systems, the inevitable strong heating can be prevented at rather long times in the so-called prethermalization regime, where the system can be analytically described by the time-independent effective model \cite{Bukov_rev,Abanin_rev,Saito2016}. 
Nevertheless,  at some particular, resonant frequencies of the drive rapid energy absorption by a Floquet system cannot be avoided. 

Recently, it has been found that the Fermi-Hubbard system resonantly absorbs energy at the series of periodic-driving frequencies \cite{Schiro2018}. Though singular behavior of the Floquet-Schrieffer-Wolff  approach at similar frequencies \cite{Schiro2018} indicates the critical role of  high-energy charge fluctuations (giving rise to the spin-exchange \cite{Spalek1977}), a clear view on the nature of the mechanism responsible for the found resonances \cite{Schiro2018} is still missing.

 \begin{figure}[b]
  \begin{center}
   \includegraphics[width=0.45\textwidth]{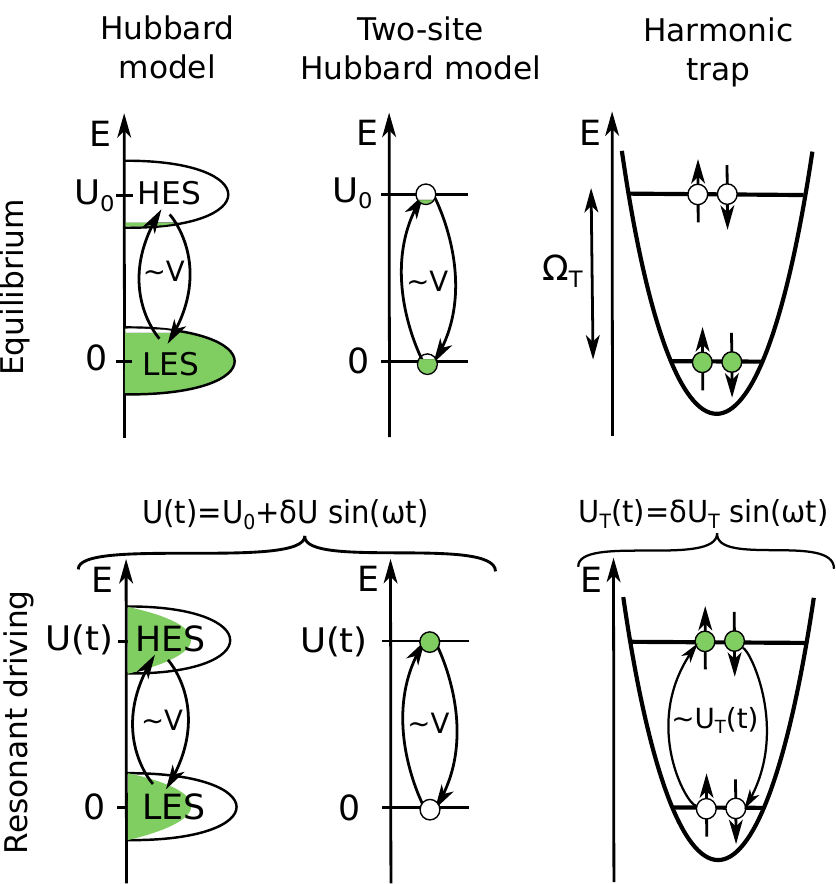}
   \caption{Schematic drawing visualizing energy scales of   Hubbard model, two-site Hubbard model and two-level harmonic trap with two fermions. 
   Green color  represents weights corresponding to the realized state of a given system: (upper panels)  equilibrium groundstate and (bottom panels)  excited state due to resonant driving. }
   \label{cartoon}\vspace{-0.3cm}
  \end{center}
 \end{figure}

Here, we shed more light on this issue by considering static, rather than nonequilibrium, perturbation expansion of the driven Fermi-Hubbard model \cite{Spalek1977,Wysokinski2017R}. 
The perturbative Hamiltonian indicates that  driving couples Hubbard bands like a classical field acts on a two-level atom. The latter system is described by the Rabi model \cite{gerry,agarwal2012},  which can support resonant rapid energy absorption at series of critical frequencies of the drive \cite{Glenn2013,Li2013,Saiko2014,Yan2016,Yan2017} in the virtually quantitative agreement with those revealed in the Fermi-Hubbard model \cite{Schiro2018}. 
The indicated correspondence to a two-level atom driven by a field is further supported by  the equivalence between dynamics of two-site Fermi-Hubbard 
and Rabi models.

Additionally, we establish that the  Rabi-resonant energy absorption is also present in a system of few fermions confined in a one dimensional harmonic trap, driven by periodically-modulated contact interactions. Such set-up is achievable in ultra-cold atoms experiments \cite{Kinoshita2004,zurn2013Pairing,zurn2012fermionization} and can be theoretically simulated with numerically exact approaches \cite{Sowinski2013,Sowinski2015Pairing, Pecak2017Ansatz,Plodzien2018,Plodzien2018b,Koscik2018,Koscik2018opty,Sowinski2019}. 

{\it Driven Hubbard model.}
Half-filled Fermi-Hubbard model on the infinitely coordinated Bethe lattice ($z\!\rightarrow\!\infty$)
\begin{equation}
 H=-\frac{V}{\sqrt{z}}\sum_{\sigma,\langle\g i\g j\rangle} c^\dagger_{\g i\sigma}c^\dagga_{\g j\sigma}+ U(t) \sum_{\g i}n_{\g i \uparrow} n_{\g i \downarrow}.
 \label{HM}
\end{equation}
with  periodically modulated interaction amplitude $U(t)=U_0+\delta U\sin(\omega t)$, and rescaled nearest-neighbor hopping $V/\sqrt{z}$, has been recently considered in Ref. \cite{Schiro2018}. It has been found that at particular driving frequencies the Fermi-Hubbard system can rapidly absorb energy. Moreover, these frequencies have been found to correlate with those equal to submultiples of $U_0$ at which the Floquet-Schrieffer-Wolff perturbation approach breaks down, indicating a critical role of the high-energy charge fluctuations.  In the following, we aim to understand their direct role. 

First, we distinguish in the Fermi-Hubbard model, in  $ U_0\gg V$  limit, the low-energy subspace (LES) without double occupancies and the high-energy subspace (HES) with double occupancies. For that purpose, the hopping operator ${T_{\g i\g j}\equiv\frac{1}{\sqrt{z}}\sum_\sigma (c_{\g i\sigma}^\dagger c_{\g j\sigma}^\dagga+ 
{\rm H.c.})}$ on the bond $\langle \g i\g j \rangle$ is split into parts acting within  LES and HES or mixing them. Mixing terms of the hopping are formally defined as $\tilde{T}^\dagga_{\g i\g j}\equiv {\rm P}_{20}T_{\g i\g j}{\rm P}_{11}$ and $\tilde{T}^\dagger_{\g i\g j}\equiv{\rm P}_{11}T_{\g i\g j}{\rm P}_{20}$ where
${\rm P}_{nm}\equiv \mathcal P_{\g i;n}\mathcal P_{\g j;m}+(1-\delta_{nm})\mathcal P_{\g i;m}\mathcal P_{\g j;n}$, $\delta_{nm}$ is the Kronecker delta and ${\mathcal P}_{\g i;n}$ is the projection operator at site $\g i$ onto the configuration with $n\in\{0,1,2\}$ electrons.
Usual nonequilibrium perturbation approach \cite{Polkovnikov2016,tSWT,Schiro2018} relies on removing the leading order mixing between subspaces, which allows to unveil system properties in the low-energy sector. However, the singularity found in the 
Floquet-Schrieffer-Wolff approach \cite{Schiro2018} suggests that such decoupling is not possible for the resonant driving, and thus system cannot be effectively described in LES. For that reason, we use static perturbation theory in small parameter $V/U_0\ll1$,
\begin{equation}
 \mathcal{H} \!=\! {\rm e}^{- S}\!H{\rm e}^{S}\!=\!  H -  [S,H]+\frac{1}{2}[S,[S,H]]  + \mathcal{O}\big(V^2/U_0^2\big)
 \label{SW}
\end{equation}
with $S=\frac{V}{U_0\sqrt{z}}\sum_{\langle \g i\g j\rangle} (\tilde{T}^\dagga_{\g i\g j}-\tilde{T}_{\g i\g j}^\dagger)$. Transformation removes the leading order mixing between subspaces only in the static part of $H$ \cite{Spalek1977,Wysokinski2017R}. As a result, the perturbative form of the driven Hubbard model up to terms of the order of $V^2/U_0^2$  spanned in LES and HES reads (note that $\delta U$ and $U_0$ can be of the same order)
\begin{equation}
\mathcal H= 
\begin{pmatrix}
  \frac{V^2}{U_0}\big(1-\frac{\delta U \sin \omega t}{U_0}\big) \sum\limits_{\langle \g i\g j\rangle }\mathbb{J}^H_{\g i\g j}&  \frac{V\delta U}{U_0} \sin \omega t \sum\limits_{\langle \g i\g j\rangle }\tilde{T}^\dagga_{\g i\g j}\\
  \frac{V\delta U}{U_0} \sin \omega t \sum\limits_{\langle \g i\g j\rangle }\tilde{T}^\dagger_{\g i\g j}   &  \frac{V^2}{U_0}\big(\frac{\delta U \sin \omega t}{U_0}-1\big)\sum\limits_{\langle \g i\g j\rangle }\mathbb{J}^L_{\g i\g j} 
 \end{pmatrix},
 \label{rabiSW}
\end{equation}
where  $\mathbb{J}^L_{\g i\g j}\equiv {\rm P}_{11}\,T^2_{\g i\g j}\,{\rm P}_{11}\simeq-4{\g S}_{\g i}\cdot{\g S}_{\g j}$  and  $\mathbb{J}^H_{\g i\g j}\equiv {\rm P}_{20}\,{T}^2_{\g i\g j}\,{\rm P}_{20}$. In that manner, Heisenberg-model description of LES   \cite{Spalek1977} is recovered in the absence of driving.

The perturbative form of the Fermi-Hubbard model \eqref{rabiSW} has a similar structure to the Rabi model. The latter in quantum optics describes a two-level atom coupled to a classical field \cite{gerry,agarwal2012} with the following Hamiltonian
\begin{equation}
 H_R=\frac{1}{2}\big(\Omega_0 +\Omega_z \sin \omega t\big)\  \sigma_z +\Omega_x\sin \omega t\   \sigma_x,
 \label{rabi}
\end{equation}
where $\sigma_{x,y,z}$ are Pauli matrices, $\Omega_0$ is level separation, and $\Omega_x$ and $\Omega_z$ are transverse and longitudinal components of a field respectively. 
Standard Rabi model, i.e. in the absence of longitudinal driving ($\Omega_z=0$) \cite{gerry,agarwal2012}, leads to the resonant population of the high energy level when field frequency is synchronized with the level separation $\omega=\Omega_0$. Additionally, for strong drivings $\Omega_x/\Omega_0\gtrsim1$, multiphoton resonances can be seen  at   
$\omega^{(n)}\approx\Omega_0/(2n+1)$, for positive integer $n$.   
In turn, when $\Omega_z\neq 0$  \cite{Glenn2013,Li2013,Saiko2014,Yan2016,Yan2017} 
resonances can appear at all integer submultiples of level separation, $\omega_{n}\approx\Omega_0/n$, $n\in\mathbb{Z}$, even for weak transverse drivings. This is due to multiphoton processes involving quanta from the longitudinal part of the driving (for details see Supplemental Material \cite{SM}).   
 
Based on the apparent similarity between Rabi and driven Fermi-Hubbard models, we propose following  interpretation of the properties of the latter \cite{Schiro2018}.
Hubbard bands, mirrored by the time-independent diagonal terms in \eqref{rabiSW}, form an effective two-level system, whereas driving couples them in a similar fashion as a classical field with transverse and longitudinal components. Multiphoton processes of the transverse driving can be disregarded, as the amplitude of the off-diagonal terms in Eq. \eqref{rabiSW} is small in comparison to the Hubbard bands splitting, $V\delta U/U_0\ll U_0$. In that manner, rapid energy absorption for frequencies close to submultiples of the separation between Hubbard bands  are associated with multiphoton processes, involving quanta from longitudinal driving.

{\it Two-site cluster.} In this part of the work we support the above interpretation to the resonantly driven Fermi-Hubbard model \cite{Schiro2018} from a different perspective.  We analyze the minimal set-up that mimics high-energy charge fluctuations, {\it i.e.} half-filled two-site Hubbard model
\begin{equation} 
 H_{2s}=-V\sum_\sigma (c_{1\sigma}^\dagger c_{2\sigma}^\dagga +{\rm H.c}) + U(t)\sum_{\g i=1}^2n_{\g i \uparrow}n_{\g i \downarrow},
 \label{2s}
\end{equation}
whose equilibrium properties are well known (see e.g. Ref. \cite{scalettar}). In turn, some of its nonequilibrium properties  have been also recently considered and experimentally realized \cite{Esslinger2017,Esslinger_arxiv}.
Due to underlying symmetries ($H_{2s}$ commutes with  total spin  and parity, exchanging $\g i=\{1,2\}$, operators) 
Hamiltonian \eqref{2s} can be block diagonalized. In result, the evolution of the system  initially prepared in the groundstate,  takes place in the subspace spanned by two spin-singlet states
\begin{equation}
\begin{gathered}
 |\psi_1\rangle=\frac{1}{\sqrt{2}}(c_{1\uparrow}^\dagger c_{2\downarrow}^\dagger -c_{1\downarrow}^\dagger c_{2\uparrow}^\dagger )|\emptyset\rangle,\\
 |\psi_2\rangle=\frac{1}{\sqrt{2}}(c_{1\uparrow}^\dagger c_{1\downarrow}^\dagger + c_{2\uparrow}^\dagger c_{2\downarrow}^\dagger )|\emptyset\rangle.
 \end{gathered}
\end{equation}
and is governed by the following  Hamiltonian 
\begin{equation}
 \mathcal{H}_{2s}=\frac{U(t)}{2}(1-  \sigma_z)- 2 V  \sigma_x.
\end{equation}
In accord with the $Z_2$ slave-spin approach \cite{Huber2009,Huber2010,Fabrizio2011},  Hamiltonian  $\mathcal{H}_{2s}$ acts in a pseudospin space, where the negative eigenvalue of $z$ component is associated with absence ($|\psi_1\rangle$) and the positive one with  presence ($|\psi_2\rangle$) of a doublon-holon pair. In fact, such structure is  reminiscent of LES and HES in the Fermi-Hubbard model  (cf. left and middle panels of Fig. \ref{cartoon}).
\begin{figure}
\begin{center}
   \includegraphics[width=0.47 \textwidth]{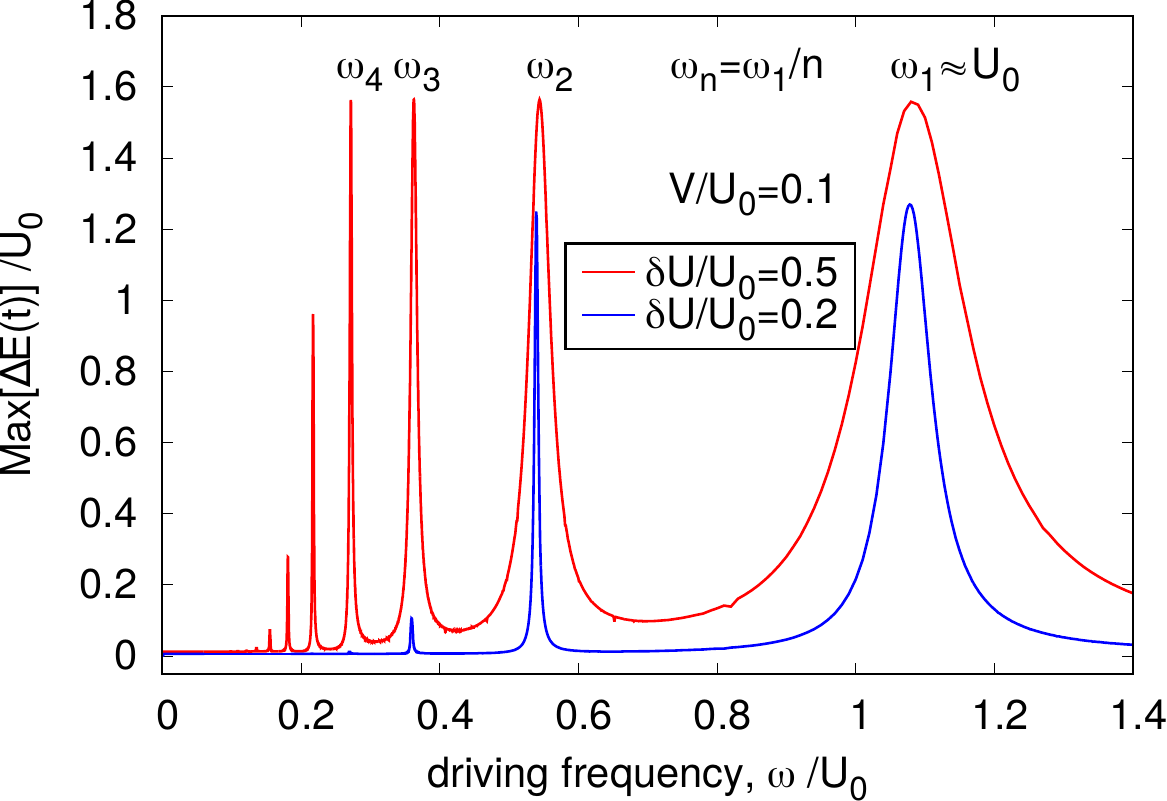}
   \caption{Maximal energy absorption from a drive Max$[\Delta E(t)]$ with respect to driving frequency $\omega$ at $t\,U_0\in[0,500]$ time scale realized by the Hubbard dimer ($\mathcal{H}_{2s}$) for $V/U_0=0.1$ and two selected driving strengths $\delta U/U_0=\{0.5,0.2\}$. Seven resonant frequencies $\omega_n= \omega_1/n$ are visible for the stronger driving while only three of them for the weaker one. Remaining resonances are realized on the longer time scale than $500$.}
   \label{emax}
  \end{center}
 \end{figure}
Hamiltonian $\mathcal{H}_{2s}$,  transformed to the basis in which  
time-independent part is diagonal, takes the form of the Rabi model (cf. Eq. \eqref{rabi})
\begin{equation}\label{H_eff_rabi2}
 \mathcal{H'}_{2s}(t) = H_R + \frac{U(t)}{2}\mathds{1},
\end{equation}
with $\Omega_0= -\sqrt{U_0^2+16V^2}$, $\Omega_z= \frac{\delta U U_0}{\Omega_0}$ and $\Omega_x= -\frac{2\delta U V}{\Omega_0}$.
In order to visualize Rabi resonances supported by the two-site Hubbard model, we
solve Schr\"odinger equation $i\partial_t|\psi(t)\rangle = \mathcal{H}_\text{2s}(t)|\psi(t)\rangle$, with initial condition $|\psi(t=0)\rangle$ being the groundstate of the non-driven system with energy $E_0$. In Fig. \ref{emax} 
we present maximal energy absorption from a drive
\begin{equation}
 \Delta E(t) =\langle\psi(t)|\mathcal{H}_{\text{2s}}(t)|\psi(t)\rangle - E_0,
 \label{ene}
\end{equation}
as a function of the driving frequency $\omega$.
Indeed, as it can be seen in Fig. \ref{emax}, the maximal-energy excited state with dominant doublon-holon pair probability can be reached for resonant driving frequencies equal to submultiples of frequency $\omega_1\simeq U_0$, {\it i.e.} $\omega_{n}\simeq \omega_1/n$, in a virtually quantitative agreement with the Fermi-Hubbard model \cite{Schiro2018}.
Our finding allows us to further study character of resonances supported by the Fermi-Hubbard model \cite{Schiro2018}  within the significantly simpler set-up of the Hubbard dimer.  

Energy scales in the Hubbard dimer follow $|\Omega_0|>|\Omega_z|\gg|\Omega_x|$ as long as $ U_0\gg V$ and $\delta U\sim U_0$. The condition $|\Omega_0|\gg|\Omega_x|$ allows to disregard multiphoton processes associated with the transverse field component \cite{Li2013,Yan2017}. Therefore, resonant rapid energy absorption at $\omega_{n>1}$ is connected to multiphoton processes involving quanta from the longitudinal driving parameterized by sizable $\Omega_z$ \cite{Li2013,Yan2017} (see Supplemental Material \cite{SM}).

{\it Few fermions in a harmonic trap.}  
In this section we show that the multiphoton processes involving quanta from transverse and longitudinal driving components are also realized by ultra-cold fermions in a one-dimensional harmonic trap with periodically modulated contact interaction. Hamiltonian for such set-up 
reads
  \begin{equation}\label{HamFerm2}
\begin{split}
 {H}(t) &= \sum_{\sigma,i} \epsilon_i a^\dagger_{\sigma i}a^\dagga_{\sigma i} + U_T(t)\sum_{ijkl}\gamma_{ijkl}a^\dagger_{\uparrow i}a^\dagger_{\downarrow j}a^\dagga_{\downarrow k}a^\dagga_{\uparrow l},
\end{split}
\end{equation}
where $a^\dagga_{\sigma i}$  ($a^\dagger_{\sigma i}$) are annihilation (creation) operators of fermion with spin  $\sigma$ in the $\varphi_i(x)$ eigenstate of the harmonic trap with frequency $\Omega_T$ fulfilling anticommutation relations $\{a_{\sigma i},a^\dagger_{\sigma j} \}=\delta_{\sigma \sigma'}\delta_{ij}$, $\epsilon_i = \Omega_T(i+1/2)$, $U_T(t)\gamma_{ijkl}$ are interaction matrix elements with $\gamma_{ijkl}   = \int\!\!\mathrm{d}x\,\varphi_i(x)\varphi_j(x)\varphi_k(x)\varphi_l(x)$, and $U_T(t)=\delta U_T\sin(\omega t)$.
Hamiltonian \eqref{HamFerm2} commutes with the particle number operator in a given spin $\sigma$, thus number of particles $N_\sigma$ is constant during time-evolution. Therefore, numerical analysis can be prepared in a subspace with fixed total particle number $N = N_\uparrow+N_\downarrow$. Next, parity of the single orbital states \big($\varphi_i(x)=(-1)^i\varphi_i(-x)$\big) results in non-zero interaction matrix elements only between Fock-states with energy difference equal to $2n\Omega_T$, with $n\in \mathbb{Z}$. 
The dynamics of the Hamiltonian \eqref{HamFerm2} is studied  
in the basis of Fock states
$|v\rangle = |n_0, n_1, \dots; m_0, m_1, \dots\rangle \sim (a_{\uparrow 0}^\dagger)^{n_0}(a_{\uparrow 1}^\dagger)^{n_1}\cdots(a_{\downarrow 0}^\dagger)^{m_0}(a_{\downarrow 1}^\dagger)^{m_1}\cdots|\emptyset\rangle$
where $|\emptyset\rangle$ is a system vacuum and $n_i, m_i \in \{0,1\}$.

The simplest set-up of two  fermions ($N_\uparrow=N_\downarrow = 1$) in two-level harmonic trap (cf. right panel of Fig. \ref{cartoon}) provides a direct reference to the Rabi physics. Matrix form of the respective Hamiltonian \eqref{HamFerm2} in the basis 
$\{ |1,0;0,1\rangle, |0,1;1,0\rangle,|1,0;1,0\rangle, |0,1;0,1\rangle \}$ is block diagonal and reads
\begin{equation}
 \begin{split}H_{2l}=
  \begin{pmatrix}
  \Omega_0 + \gamma_2 U_T & \gamma_2 U_T & 0 &0\\
     \gamma_2 U_T& \Omega_0 + \gamma_2 U_T  &0 & 0\\
0 & 0 & \gamma_0 U_T &  \gamma_2 U_T\\
0 & 0 & \gamma_2 U_T&  2\Omega_0 +\gamma_1 U_T
  \end{pmatrix}.
 \end{split}
\end{equation}
Calculated non-zero integrals of harmonic oscillator eigenstates  read $\gamma_0=\gamma_{0000}\simeq0.4$, $\gamma_1=\gamma_{1111}\simeq0.3$, $\gamma_2\!\equiv\!\gamma_{1010}\!=\!\gamma_{0101}\!=\!\gamma_{0110}\!=\!\gamma_{0011}\!=\!\gamma_{1001}\!=\!\gamma_{1100}\!\simeq\! 0.2$. 
Given that an initial state at $t=0$ is the non-interacting ground state,  dynamics  is restricted to the space spanned by  
$|1,0;1,0\rangle=a_{\uparrow0}^\dagger a_{\downarrow0}^\dagger|\emptyset\rangle$ and $|0,1;0,1\rangle= a_{\uparrow1}^\dagger a_{\downarrow1}^\dagger|\emptyset\rangle$ states. 
Therefore, we can describe  time-evolution of the system with the simple two-level Hamiltonian which is equivalent to the Rabi model (cf.  Eq. \eqref{rabi}) 
 \begin{equation}\label{H2l_rabi}
H_{2l}'\equiv H_R +\Big(\Omega_T+\frac{\gamma_1+\gamma_0}{2}U_T(t)\Big) \mathds{1} 
\end{equation}
 with $\Omega_0=-2\Omega_T$, $\Omega_z=(\gamma_1-\gamma_0)\delta U_T$ and $\Omega_x=\gamma_2\delta U_T$. Here, when $\delta U_T\lesssim \Omega_T$, energy scales follow $\Omega_0\gg \Omega_x> \Omega_z$. These relations suggest that multiphoton processes involving quanta only from transverse driving can be disregarded.     

  \begin{figure}[t]
  \begin{center}
   \includegraphics[width=0.47\textwidth]{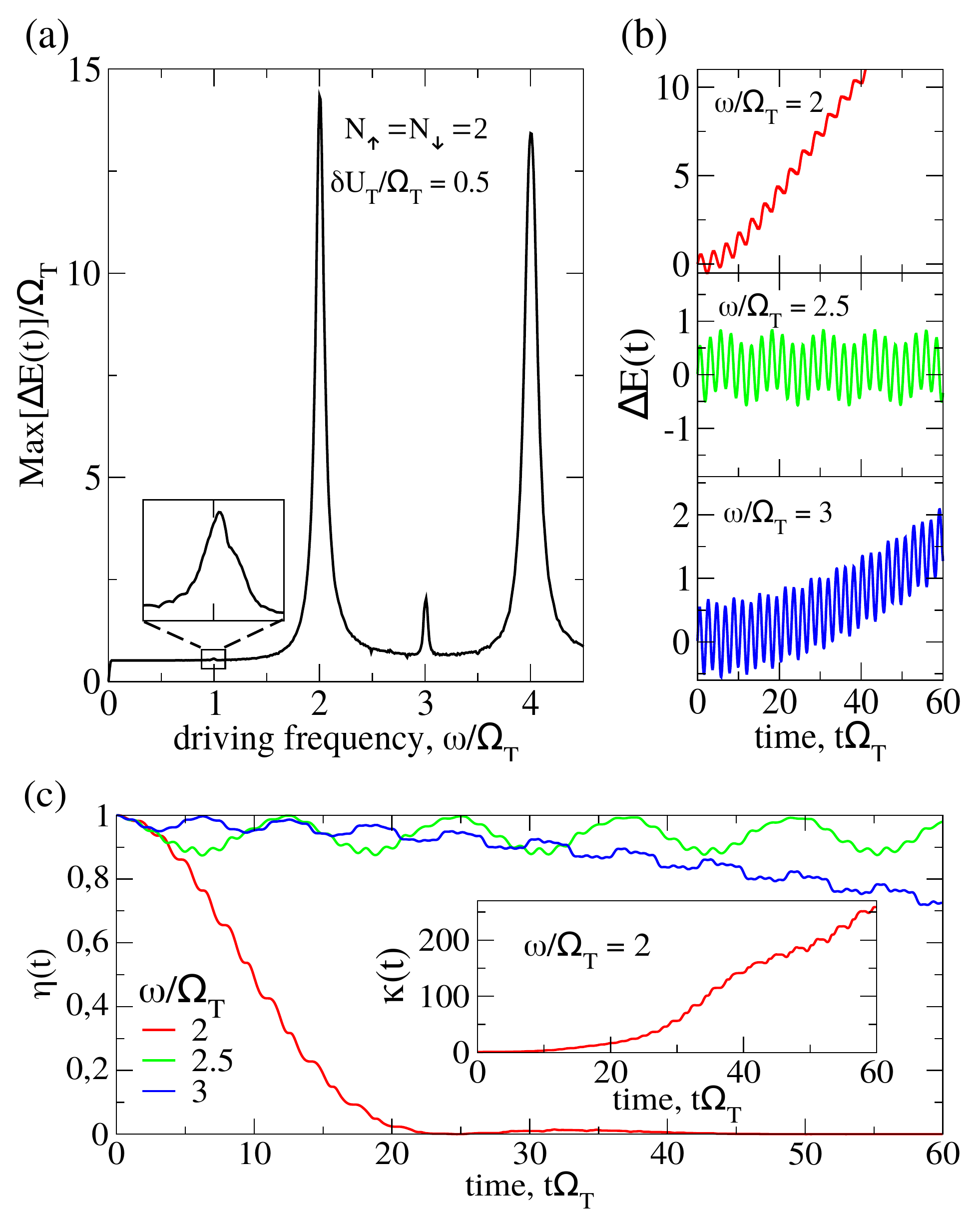}
   \caption{Dynamic properties of four fermions ($N_\uparrow=N_\downarrow=2$) in a harmonic  trap driven by  periodically modulated contact interaction:  (a) maximal energy absorbed from the drive Max$[\Delta E(t)]$   as a function of driving frequency, $\omega$ for $t\,\Omega_T\leq60$; (b) real-time evolution of $\Delta E(t)$ on resonance ($\Omega=\{2\Omega_T, 3\Omega_T\}$) and off-resonance ($\Omega=2.5\Omega_T$); (c) Real-time evolution of Lochschmidt echo $\eta(t)$ for the same driving frequencies. Inset of (c) shows on-resonance inverse participation ratio $\kappa(t)$ (see main text for the definition).}
   \label{trap}
  \end{center}
 \end{figure}

In the following, within the numerically exact approach \cite{Haugset1998,Plodzien2018} we consider time evolution of a richer system, i.e. four fermions $N_\downarrow=N_\uparrow=2$  (for discussion on numerical accuracy see Supplemental Material \cite{SM}).
In Fig. \ref{trap}(a) we present maximal energy absorption of the system as a function of driving frequency (defined in the same manner as in Eq.~\eqref{ene}) during times $t\Omega_T<60$. One can distinguish four resonances at $\omega\simeq\{1,2,3,4\}\times\Omega_T$. Peaks at $\omega=2n\Omega_T$ can be associated with the usual Rabi-resonances when the frequency of the transverse driving is synchronized to the separation between levels ($\omega=2\Omega_T$). On the other hand, in Fig. \ref{trap}(a) at $\omega\simeq(2n-1)\Omega_T$ one can distinguish resonances due to multiphoton processes involving quanta also from the longitudinal driving. These are less pronounced at accessible time scales because amplitude of longitudinal driving  constitutes the smallest energy scale in the problem (cf. Eq. \eqref{H2l_rabi}). For clarity in  Fig. \ref{trap}(b) we present real-time evolution of absorbed energy for two resonant (at $\omega2\Omega_T, 3\Omega_T$)  and off-resonant (at $\omega=2.5\Omega_T$) frequencies of the drive.
In order to characterize the wave function of the system during time-evolution, in Fig. \eqref{trap}(c) we present Lochschmidt echo, $\eta(t)=\big|\langle \psi_0|\psi(t)\rangle|^2$, \textit{i.e.} overlap between time-evolved and initial state. 
For off-resonant driving the wave function   does not evolve far from the initial state, as $\eta$ stays near unity. Contrary, for the resonant frequencies of the drive, $\eta$ escapes from the unity and  for $\omega=2\Omega_T$ rapidly reaches $0$. In the inset of Fig. \ref{trap}(c) we further characterize  structure of the wave function in time for the  resonant driving $\omega=2\Omega_T$ in terms of the inverse participation ratio, $\kappa(t)^{-1}=\sum_{v} |\langle v|\psi(t)\rangle|^4$.  This coefficient constitutes a good estimation of the number of Fock states with the dominant weight in a wave function. In particular, when a wave function is equally distributed over all possible basis states, $\kappa$ reduces exactly to the dimension of the Hilbert space, i.e. $\kappa=\rm{dim}{H}$.
In the considered situation of resonantly driven trapped fermions, rapidly increasing  $\kappa$ indicates that the wave function is quickly smeared out over large number of Fock states. 
  
{\it Summary.} In the present work, we study the mechanism of recently found resonant energy absorption in the periodically interaction-driven Fermi-Hubbard system at series of driving frequencies \cite{Schiro2018}. 
With the help of the static perturbation approach we argue that  Hubbard bands are 
coupled by driving in a similar manner as a classical field with transverse and longitudinal components acts on a two-level atom. 
The latter scenario, formally described by the quantum optical Rabi model, though significantly simpler than the Fermi-Hubbard system, supports resonant rapid energy absorption \cite{Glenn2013,Li2013,Saiko2014,Yan2016,Yan2017} at virtually the same frequencies of a drive. 
Our interpretation is further supported by equivalency between dynamics of driven two-site half-filled Hubbard and Rabi models.
Relying on this insight, we show that few contact-interaction-driven fermions confined in a one dimensional harmonic trap display the same resonant behavior of the Rabi model with a field with transverse and longitudinal components.

 \begin{figure*}[t]
  \begin{center}  
  \vspace{-0.4cm}
    \includegraphics[width=0.9\textwidth]{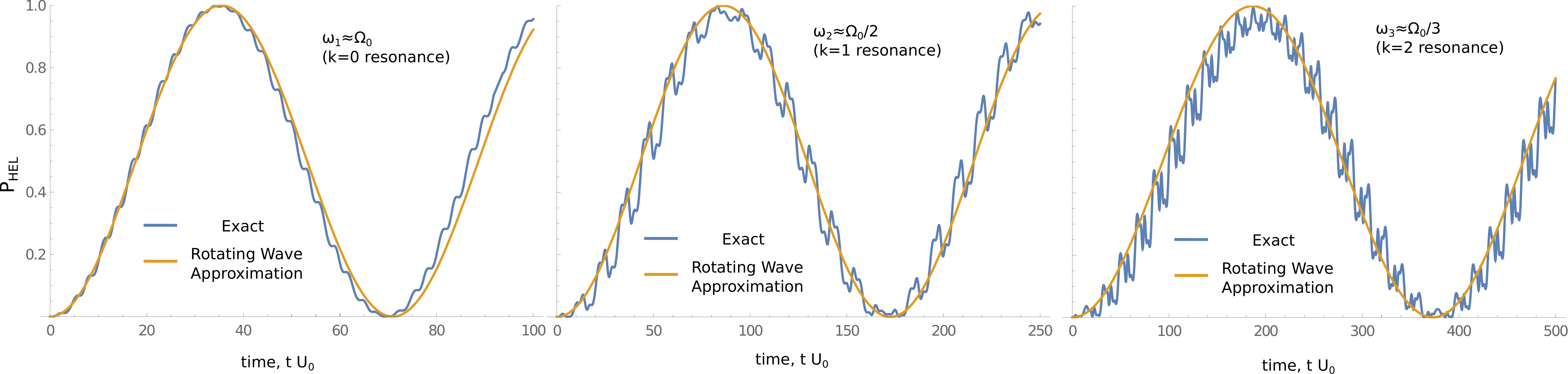}
  \vspace{-0.4	cm}
  \end{center} 
 \caption{Comparison between time-evolution of exact resonant  population of HEL in Hubbard dimer (Eq.~(8) in the main manuscript) and RWA estimate.}
\label{fig2s} 
\end{figure*}

  {\it Acknowledgments.}  
Discussions with  Wojciech Brzezicki, Arkadiusz Kosior, \L{}ukasz Cywi\'nski,  Tomasz Sowi\'nski and Michele Fabrizio are greatly acknowledged.
The  International Centre for Interfacing Magnetism and Superconductivity with
Topological  Matter  project  is  carried  out  within  the  International  Research  Agendas  program  of  the  Foundation  for Polish Science co-financed by the European Union under the European Regional Development Fund.  This work was partially supported by the (Polish) National Science Center, grant No. 2016/22/E/ST2/00555 (MP).

\vspace{0.5cm}

 \begin{center}
\Large{Supplemental Material}\end{center}

\appendix
 \section{Multiphoton resonances in Rabi model}

Rabi model in the presence of a field with both transverse and longitudinal components  (cf. Eq. (4) in the main manuscript) reads:
\begin{equation}
 H_R=\frac{1}{2}\big(\Omega_0 +\Omega_z \sin \omega_z t\big)\  \sigma_z +\Omega_x\sin \omega_x t\   \sigma_x,
\end{equation}
where $\omega_x=\omega_z=\omega$. Subscripts for frequencies of a field in each of its components are artificially added for the purposes of clarity of forthcoming discussion.

\subsection{$\Omega_z=0$}
In the absence of the longitudinal field component, $\Omega_z=0$ when the driving frequency is synchronized with the
level separation, i.e. $\omega_x=\Omega_0$ the model supports resonant (oscillating) population of the high energy level (HEL). Then, if initially HEL is empty, its population in time for the resonant transverse driving  can be approximated quite well by 
\begin{equation}
P_{HEL}(t)=\sin^2 (\Omega_R t/2) 
\label{hel}
\end{equation}
 where $\Omega_R=\Omega_x$ is called the Rabi frequency. The Rotating Wave Approximation (RWA), under which above result is obtained, ignores the so-called counter-rotating terms of the transverse field in model (1) \cite{Yan2017}. These in turn can be important in the strong driving regime, $\Omega_x/\Omega_0\gtrsim1$ and can lead to multiphoton resonances at $\omega_x^{(p)}=\Omega_0/(2p+1)$,
where $p$ is a positive integer.

\subsection{$\Omega_z\neq0$} 
Non-zero longitudinal driving ($\Omega_z\neq0$) leads to a different  structure of resonances. 
Both transverse and longitudinal drivings can be considered as separate fields, and thus processes involving quanta from both of them can lead to multiphoton resonances. In the Hubbard dimer (cf. Eq. (8) in the main manuscript) the transverse driving amplitude is smaller than the level separation
\begin{equation}
 |\Omega_x/\Omega_0|=\frac{2\delta U V}{U_0^2+16V^2}\ll 1
\end{equation}
because $\delta U\lesssim U_0$ and $V\ll U_0$.
This follows that counter-rotating terms, and thus multiphoton processes associated with the transverse field only can be disregarded. The application of RWA to the problem, validated in that manner, reveals the existence of  multiphoton resonances due to additional quanta from longitudinal field, {\it i.e.} at $\Omega_0=\omega_x + k\omega_z$, with $k\in \mathbb{Z}$ \cite{Li2013,Yan2017}. As a result, the oscillating population of HEL at resonances can be approximated by formula \eqref{hel} with the effective Rabi frequency \cite{Li2013},
\begin{equation}
 \Omega^{\rm eff}_R=\big|\Omega_xJ_k\big(\Omega_z/\omega_z\big)\big|,
\end{equation}
where $J_k$ is a Bessel function of $k$-th order.  Effective Rabi frequency can be further simplified by employing condition for multiphoton resonance and $\omega_x=\omega_z=\omega$,
\begin{equation}
 \Omega^{\rm eff}_R=|\Omega_xJ_k ( \Omega_z (k+1)/\Omega_0 )|.
\end{equation}
In Fig. \ref{fig2s} we plot exact numerical population of HEL for a driven Hubbard dimer with $\delta U=0.5$ (Eq.~(8) in the main manuscript) for first three resonances $\omega = \omega_{\{1-3\}}$ (cf. Fig. 2 in the main manuscript), together with estimate provided by RWA for a subsequent $k=\{0-2\}$.

The above analysis underlines predominant role of the longitudinal field component, generated by the modulation of interaction in the Hubbard dimer, for the formation of resonances at submultiples of $U_0$.

\section{Convergence test for interaction-driven fermions in a harmonic trap}
\begin{figure}[h]
  \begin{center}  
    \includegraphics[width= 0.45 \textwidth]{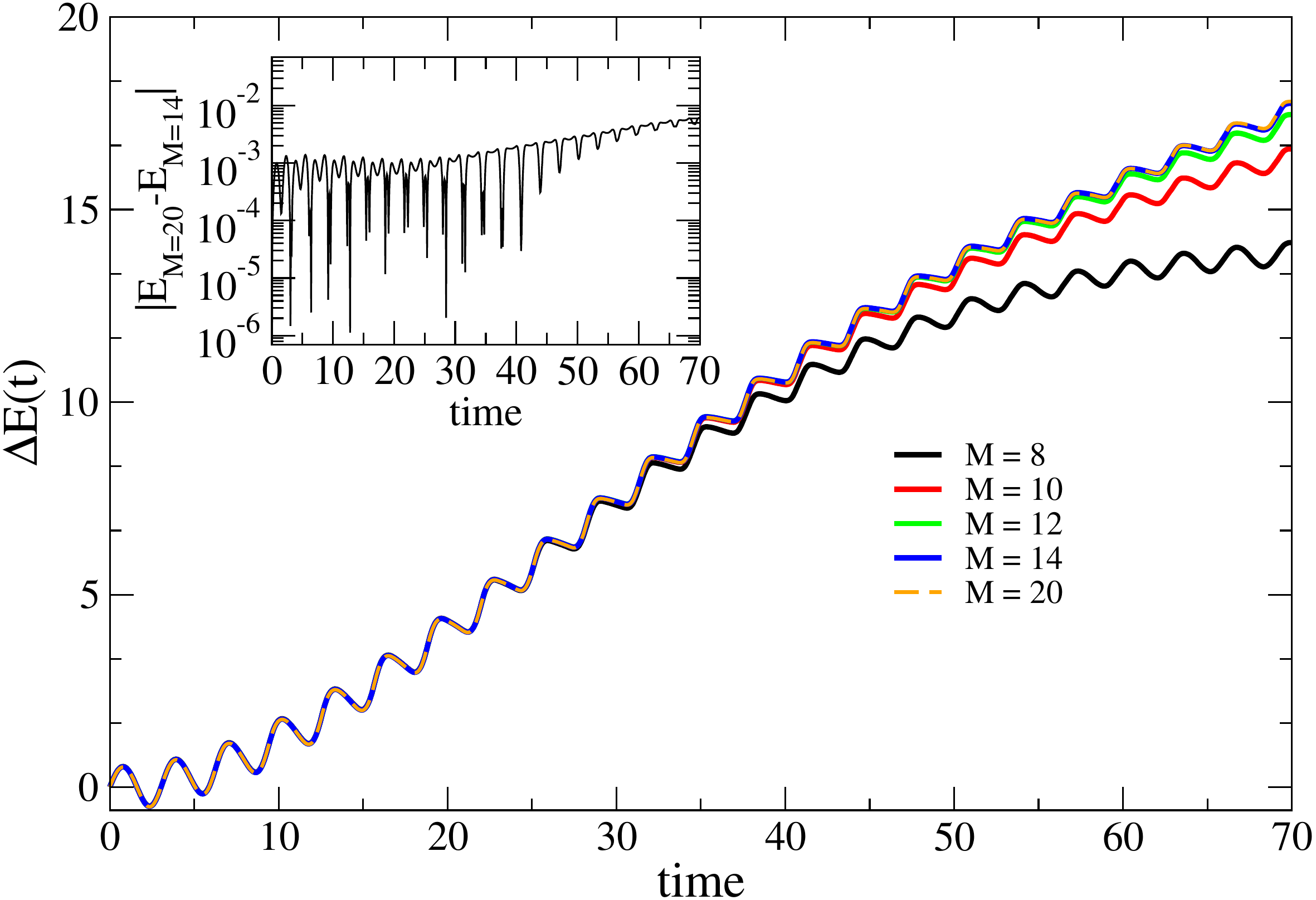}
  \vspace{-0.4	cm}
  \end{center} 
 \caption{Time evolution of the resonant ($\omega = 2\Omega_T$) energy absorption for $\delta U_T = 0.5$. Initial state is a non-interacting ground state of four fermions. Convergence is obtained for cutoff of the number of harmonic oscillator eigenstates $M\ge 14$. 
 Inset presents absolute value of the difference between energy evolution with cutoff $M = 20$ and $M = 14$.}
\label{fig0} 
\end{figure}

In this section we present numerical test for convergence of the dynamics of fermions in a one-dimensional harmonic trap driven by time-modulation of interactions $U_T(t) = \delta U_T\sin(\omega t)$. Hamiltonian for the two-component fermionic mixture 
\begin{equation}
\begin{split}
 {H}(t) &= \sum_{\sigma,i} \epsilon_i a^\dagger_{\sigma i}a^\dagga_{\sigma i} + U_T(t)\sum_{ijkl}\gamma_{ijkl}a^\dagger_{\uparrow i}a^\dagger_{\downarrow j}a^\dagga_{\downarrow k}a^\dagga_{\uparrow l},
\end{split}
\end{equation}
is expandend in the basis of Fock states 
$|v\rangle = |n_0, n_1, \dots; m_0, m_1, \dots\rangle \sim (a_{\uparrow 0}^\dagger)^{n_0}(a_{\uparrow 1}^\dagger)^{n_1}\cdots(a_{\downarrow 0}^\dagger)^{m_0}(a_{\downarrow 1}^\dagger)^{m_1}\cdots|\emptyset\rangle$,
where $|\emptyset\rangle$ is a system vacuum, $n_i, m_i \in \{0,1\}$, and $i$ enumerates first $M$ eigensates of a harmonic trap, \textit{i.e.} $ i \in [0,M)$.
In the following we consider  initial state of the system as a non-interacting groundstate of four fermions $N_\downarrow = N_\uparrow = 2$.
Figure \ref{fig0} presents evolution of absorbed energy  from the drive fo resonant driving frequency $\omega = 2$ and amplitude $\delta U_T = 0.5$ ($\Omega_T=1$) for different cutoffs of the number of  harmonic oscillator eigenstates $M = \{8, 10, 12, 14, 20 \}$.  One can observe convergent dynamics for $M \ge 14$ eigenstates.

  
%

\end{document}